\documentclass[usenatbib]{mnras}

\usepackage{newtxtext,newtxmath}
\usepackage{lipsum}

\usepackage[T1]{fontenc}
\usepackage{ae,aecompl}
\usepackage{xspace}

\usepackage{graphicx, subfigure}	
\usepackage{amsmath}	
\usepackage{amssymb}	
\usepackage{color}

\graphicspath{{./figs/}}
\newcommand{\beq}{\begin{equation}}
\newcommand{\eeq}{\end{equation}}
\newcommand{\beqry}{\begin{eqnarray}}
\newcommand{\eeqry}{\end{eqnarray}}
\newcommand{\beqrys}{\begin{subequations}\begin{eqnarray}}
\newcommand{\eeqrys}{\end{eqnarray}\end{subequations}}

\newcommand{\milc}{{\rm MILC}\xspace}

\newcommand{\Planck}{{\it Planck}\xspace}
\newcommand{\WMAP}{{\it WMAP}\xspace}

\def\eq#1{{Eq.~(\ref{#1})}}

\def\sec#1{{Sec.~\ref{#1}}}
\def\fig#1{{Fig.~\ref{#1}}}
\def\sec#1{{Sec.~\ref{#1}}}

\newcommand{\update}[1]{{\textcolor{black}{#1}}}
\newcommand{\JCparagraph}[1]{{\vspace{1mm}\noindent {\it #1}}}

\title
[The Moment ILC method]
{Combining ILC and moment expansion techniques for extracting average-sky signals and CMB anisotropies}

\author[Rotti \& Chluba]{Aditya Rotti\thanks{E-mail:aditya.rotti@manchester.ac.uk} and Jens Chluba\thanks{E-mail:jens.chluba@manchester.ac.uk}
\\
Jodrell Bank Centre for Astrophysics, School of Physics and Astronomy, The University of Manchester, Manchester M13 9PL, U.K.
}

\date{\vspace{-0mm}{Accepted 2019 --. Received 2019 --}}

\pubyear{2019}

\begin{document}

\maketitle

\begin{abstract}
The method of weighted addition of multi-frequency maps, more commonly referred to as {\it Internal Linear Combination} (ILC), has been extensively employed in the measurement of cosmic microwave background (CMB) anisotropies and its secondaries along with similar application in 21cm data analysis. \update{Here we argue and demonstrate that ILC methods can also be applied to data from absolutely-calibrated CMB experiments to extract average-sky signals in addition to the conventional CMB anisotropies.} The performance of the simple ILC method is, however, limited, but can be significantly improved by adding constraints informed by physics and existing empirical information. In recent work, a moment description has been introduced as a technique of carrying out high precision modeling of foregrounds in the presence of inevitable averaging effects. We combine these two approaches to construct a heavily constrained form of the ILC, dubbed \milc, which can be used to recover tiny \update{monopolar} spectral distortion signals in the presence of realistic foregrounds and instrumental noise. This is a first demonstration for measurements of the monopolar and anisotropic spectral distortion signals using ILC and extended moment methods. We also show that CMB anisotropy measurements can be improved, reducing foreground biases and signal uncertainties when using the \milc. While here we focus on CMB spectral distortions, the scope extends to the 21cm monopole signal and $B$-mode analysis. We briefly discuss augmentations that need further study to reach the full potential of the method. 
\end{abstract}



\section{Introduction}
The anisotropies of the cosmic microwave background (CMB), beyond doubt, have greatly helped in establishing the standard $\Lambda$CDM concordance model, with the key cosmological parameters being known to percent-level precision or better \citep{WMAP_params, Planck2013params, Planck2015params}. To reach this unprecedented precision, many obstacles had to be overcome, including an extreme control of systematic, calibration uncertainties and foreground separation \citep{Planck2016HFIsys, Planck2016Cell, Planck2016fore}.
%
To mitigate foregrounds, many independent methods have been developed \citep[e.g.,][]{Tegmark2003_ilc, SMICA2003, Eriksen_2008, Remazeilles2011a}. One of them is the {\it Interal Linear Combination} \citep[ILC, ][]{Tegmark2003_ilc}, which has proven highly successful in extracting foreground-cleaned CMB maps with current \citep[e.g.,][]{Planck2013components, Planck2016ymap, Planck2016fore} and future experiments \citep[e.g.][]{Remazeilles2016, Remazeilles2018}. 

In this work, we are mainly interested in studying spectral distortions of the CMB. While the CMB energy spectrum has been shown to be extremely close to that of a blackbody \citep{Mather1994, Fixsen1996}, minor deviations from this spectrum are expected even within standard $\Lambda$CDM \citep[e.g.,][]{Chluba2016}. The largest spectral distortions are present in the monopole sky. 
These signals are created through energy exchange and photon production in the early phases of cosmic history \citep{Zeldovich1969, Sunyaev1970mu, Illarionov1974, Burigana1991, Hu1993, Sunyaev2009, Chluba2011therm, Chluba2015GreensII}. A distortion dipole is furthermore induced due to our motion with respect to the CMB restframe, very much like the CMB temperature dipole \citep{Danese1981, Balashev2015, Burigana2018CORE}.

Spectral distortion require absolutely-calibrated measurements of the CMB spectrum. Building on the heritage of {\it COBE/FIRAS}, this can be achieved with instrument concepts similar to {\it PIXIE} \citep{Kogut2011PIXIE, Kogut2016SPIE, Kogut2019WP}.
This could allow constraining many standard and non-standard processes occuring in the early Universe, at phases inaccessible by any other means \citep[for recent overview see][]{Chluba2019Voyage}.
While ILC techniques have been used to measure components with spatial anisotropies, so far they have not been applied to average-sky (i.e., monopolar) signals, a generalization we explore here. 

In this paper, we will demonstrate that ILC methods can be directly applied to absolutely-calibrated maps, allowing an extraction of the average-sky signal. The simple (i.e., blind) ILC method is, however, limited and has to be augmented by moment expansion methods \citep{Chluba2017foregrounds} to capture the inevitable foreground averaging effects, yielding the {\it Moment ILC} method (\milc, see Sect.~\ref{sec:methods}). Without this extension, large biases and enhanced uncertainties in the recovered signals remain (see Fig.~\ref{fig:map_stat}). This conclusion extends to anisotropic signals, as we demonstrate here (see Fig.~\ref{fig:sig_rec_fn_of_mom}). 

We focus on a {\it proof-of-concept} study, demonstrating the main aspects of the \milc and how it can be applied. Further studies are required for providing concrete CMB distortion forecasts, extending first studies \citep{Vince2015, Mayuri2015, abitbol_pixie}, which largely neglected spatial information. \update{These first studies are simplistic and the demonstrations we make in this work will pave the way for more realistic forecasts in the near future.} \update{In addition,} we can already anticipate that the \milc method can be applied more broadly, e.g., to the extraction of global 21cm signals \citep[e.g.][]{Pritchard2010} and CMB polarization $B$-modes, as we briefly discuss.
\vspace{-4mm}
\section{Methods}
\label{sec:methods}
Before diving into details of the analysis, we provide a brief introduction to the methods we employ in this work. In particular we present some details of standard ILC methods in \sec{sec:ilc} and the preliminaries of the methods of moment modeling of foregrounds in \sec{sec:moments} needed for the \milc.

\vspace{-2mm}
\subsection{Standard ILC methods}
\label{sec:ilc}
ILC methods have been extensively used in the analysis on multi-frequency microwave maps originally to extract maps of the CMB temperature and polarization anisotropies \citep{Tegmark2003_ilc, PlanckCollaboration2015_diffusecmb} and more recently for the measurement of $y$-distortions \citep{Planck2013ymap, Remazeilles2019}.
The multi-frequency maps, $d_{\nu i}$, can be expressed as,
\beq
d_{\nu i } = \sum_c s^c_{\nu} \tau_i^{c}  + n_{ \nu i}\, ,
\eeq
where $\nu$ denotes the observing frequency and index `$i$' denotes the pixel index; $\tau^c$ denote spatial map of the component `c', which has the spectral energy distribution (SED) $s^c_{\nu}$; $n$ denotes the measurement noise. Note that this equation is not written in any particular basis and hence the index `$i$' can either be interpreted as the address of a pixel in real or in harmonic space.

Using some prior information on the spectral components of the data, and given that the data is measured with sufficient frequency sampling, we can solve for the map of each component of interest as follows:
\begin{align}
\hat{\update{\tau}}^{c_0}_i &= \sum_{\nu} w^{c_0}_{\nu } d_{\nu i } = w^{T}_{c_0} \cdot d  \,.
\label{eq:weighted_linear_combo}
\end{align}
This can be achieved by constructing suitable weights $w^{c_0}_{\nu}$ such that, they have unit response to the SED of the component of interest: $ \sum_{\nu} w^{c_0}_{\nu } s^{c_0}_{\nu} = w^{T}_{c_0} \cdot s_{c_0}=1$. Note that we have introduced the hat notation to distinguish the separated component map $\hat{\update{\tau}}^{c}$ from the true component map $\update{\tau}^{c}$. On using these weights it is easy to see that \eq{eq:weighted_linear_combo} reduces to the following form,
\beq
\hat{\update{\tau}}^{c_0} = \update{\tau}^{c_0} + \sum_{c}^{c \neq c_0} \left( w^{T}_c \cdot s^c \update{\tau}^{c} + w^{T}_c n \right) = \update{\tau}^{c_0} + \mathcal{B}^{c_0} + n^{c_0} \,,
\eeq
where we have suppressed the spectral and spatial indices for brevity. It is important to note that the solution generally has an additive bias $\mathcal{B}^{c_0}$, and one can understand this as originating from the existence of non-zero projections of the SED of $c_0$ on the SEDs of other spectral components of the map. It is also important to note that the excess variance on the reconstructed map has contributions from the bias, the measurement noise and the chance correlation between the two. Therefore it is important to realize that an assessment on the precision of the reconstructed map necessarily requires a thorough understanding of the interplay between the additive bias and the variance in the map and their relative amplitudes.

One can optimize the solution of the component map by demanding that the weights must minimize the variance of the reconstructed component map, leading to the standard ILC solution \citep{Tegmark2003_ilc}.
One can further optimize the solution by demanding that the weights in addition to having unit response to $s^{c_0}$ and minimizing the variance, must simultaneously show zero response to the SEDs of selected components in the map. This can be concisely written using a vector matrix formulation as follows,
 \begin{align}
 w^T_{c_0} \cdot [s^{c_0},s^{c_1},s^{c_2} \cdots s^{c_n}]&=[1,0,0 \cdots 0]  
 \,, \\
 \nonumber
 \leftrightarrow 
 \qquad 
 w^T_{c_0} V &= e^{T}_{c_0} \,.
 \end{align}
This generalized minimization problem can be solved using the method of Lagrange multipliers and it can be shown that the weights are given by the following compact expression,
\begin{align}
w^{T}_{c_0} &= e^{T}_{c_0} [V^T \mathcal{C}^{-1} V]^{-1} V^T \mathcal{C}^{-1} \,, \label{eq:cilc_old}
 \end{align}
where $\mathcal{C}$ denotes the data covariance matrix and all other symbols have the same meaning as before. Since these are matrix operators it is important to note that the order in which the elements appear is critical. Equation~\eqref{eq:cilc_old} presents a more general form of the ILC referred to as the constrained ILC (cILC) and was introduced in \citet{Remazeilles2011a} and has been used to construct Sunyaev-Zeldovich (SZ) free CMB maps and vice versa. More recently this method has been shown to allow measuring the relativistic electron temperature of galaxy clusters, introducing {\it semi-blind}\footnote{Here, semi-blind means that partial constraints on the dust have been used, admitting that we only have incomplete knowledge of the true dust SED.} constraints on the dust SED \citep{Remazeilles2019}.

The general solution for the weights given in \eq{eq:cilc_old} reduces to the simple ILC when $V = s^{c_0}$ and $e_{c_0}=1$. The solution written in the form given in \eq{eq:cilc_old}, gives the impression that it is only possible to solve for one component at a time. Presenting the solution in the following form,
\begin{subequations}
\begin{align}
\label{eq:cilc}
[\hat{\update{\tau}}^{c_0},\hat{\update{\tau}}^{c_1},\hat{\update{\tau}}^{c_2} \cdots \hat{\update{\tau}}^{c_n}] &= [V^T \mathcal{C}^{-1} V] \, V^T \mathcal{C}^{-1} d \,,  \\
\leftrightarrow 
 \qquad 
\hat{\vec{\update{\tau}}} &= w^T d\,,
\end{align}
\end{subequations}
makes it clear that the filter actually returns a vector $\hat{a}$ of minimum variance maps with mutually orthogonal SEDs, the projection operator being defined with respect to the data covariance matrix. 


\vspace{-3mm}
\subsection{Moment modeling of the observed sky}
\label{sec:moments}

The moment method can be used to efficiently model the effect of averaging (i.e., along the line of sight, in the beam and averaging operations in the data processing) for known fundamental SEDs \citep{Stolyarov2005,Chluba2017foregrounds}. The limitations in our ability to model the foregrounds to arbitrary precision is a consequence of the limited sensitivity and frequency coverage of observations and, additionally, our ignorance about the presence of totally `new' foregrounds components, which one may discover on making more refined measurements\footnote{The moments of currently known foreground might span a broader set of SEDs but may not be optimal for these new components. So this can also be cast as large increase in sensitivity and frequency coverage, if we wanted to model new foregrounds into the existing set of SED basis functions.}. In this section, we briefly introduce the basic idea and refer the reader to \citet{Chluba2017foregrounds} for a more detailed and discussions specific to certain foregrounds (dust, synchrotron etc.) \update{and the broader scope of the method}. 

Let us assume that we know that the emission from each volume element along a given line of sight is described by a SED $s_{\nu}(\bf{p})$ with some known functional form, parameterized by a vector of parameters: ${\bf p} = [p_0, p_1, p_2 \cdots p_n]$. Assuming there are multiple elements emitting along a given line of sight, $\hat{n}$, the net observed intensity from that direction is merely a sum of emission from each of the elements,
\beq
I_{\nu}(\hat{n}) =  \int A(\hat{n}, l)\, s_{\nu}( {\bf p} (\hat{n}, l)) \,{\rm d}l \,, \label{eq:add_emm_los}
\eeq
where ${\bf p}_l (\hat{n})$ denotes that the parameters characterizing (defining the spectral shape) the emission change along $\hat{n}$ and $A_l$ sets the overall amplitude of the emission. It is very likely that the emission from the different elements is characterized by different vectors ${\bf p}$. Given that there are great number of emitting elements, one may, equivalently, construct a statistical model for the observed intensity, in which \eq{eq:add_emm_los} can be re-cast in the following form,
\beq
I_{\nu}(\hat{n}) = A_{\nu_0 }(\hat{n}) \! \int s_{\nu}( {\bf p}) \,\mathcal{P}(\hat{n},{\bf p}) \,{\rm d}{\bf p}\,, \label{eq:statistical_sed_mod}
\eeq
where $\mathcal{P}(\hat{n},{\bf p})$ denotes the multi-dimensional probability distribution function of the component of ${\bf p}$, and the parameter $A_{\nu_0}(\hat{n})$ is amplitude of the observed SED at some pivot frequency $\nu_0$, introduced to allow for modeling the overall amplitude of the SED. In practice we never have access to this probability distribution, but we have a reasonable idea about the variety of SEDs contributing to the observed sky. One can now Taylor expand the emission law $s_{\nu}$ around some pivot parameter vector $\bar{{\bf p}}$,
\begin{align}
I_{\nu}(\hat{n}) &= A_{\nu_0 }(\hat{n}) \!\int \mathcal{P}(\hat{n}, {\bf p}) \,{\rm d}{\bf p} \,\Bigg\lbrace s_{\nu}( \bar{{\bf p}}) \\
&\;+ \sum_i \frac{(p_i - \bar{p}_i)}{1 !} \frac{\partial s_{\nu}}{ \partial p_i} \Big|_{\bar{\bf p}} + \sum_{i,j} \frac{(p_i - \bar{p}_i)(p_j - \bar{p}_j)}{2 !} \frac{\partial^2 s_{\nu}}{ \partial p_i \partial p_j} \Big|_{\bar{\bf p}}  \nonumber \\
&\quad+ \sum_{i,j,k} \frac{(p_i - \bar{p}_i)(p_j - \bar{p}_j)(p_k - \bar{p}_k)}{3 !} \frac{\partial^2 s_{\nu}}{ \partial p_i \partial p_j \partial p_k} \Big|_{\bar{\bf p}} + \cdots \Bigg\rbrace \nonumber \,.
\end{align}
Since the SED and its derivatives are computed at the fixed pivot parameters $\bar{\bf p}$, these spectral functions are constants and can be pulled out\footnote{We neglect the effect of frequency-dependent beams, which can complicate this step further \citep{Chluba2017foregrounds}.} of the integral and each of the integrals over the PDF can be understood as the corresponding amplitude weighted moment $\eta$ of the distribution. This allows us to model the observed intensity in the language of the moments of the parameter distribution,
\begin{align}
\label{eq:mom_sed_mod}
I_{\nu}(\hat{n}) &= A_{\nu_0} (\hat{n}) \,s_{\nu}(\bar{ \bf p}) + \sum_{i} \eta_i (\hat{n})\,  s^{i}_{\nu}(\bar{\bf p}) \\
&\quad
+\frac{1}{2 !}\sum_{ij} \eta_{ij} (\hat{n})\,  s^{ij}_{\nu}(\bar{\bf p}) + \frac{1}{3 !}\sum_{ijk} \eta_{ijk} (\hat{n})\,  s^{ijk}_{\nu}(\bar{\bf p}) + \cdots \nonumber \,,
\end{align}
where $\eta$ denotes the `amplitude-weighted moments' of the of the parameter distribution characterizing the total SED.
Here, it is important to mention that since the derivative operator with respect to different parameters commute: $s^{ij}_{\nu} \equiv s^{ji}_{\nu}, s^{ijk}_{\nu} \equiv s^{ikj}_{\nu}, ...$, one needs to appropriately take care of these degenerate SED vectors.
One of the beautiful aspect of \eq{eq:mom_sed_mod} is that the spectral and spatial parts are written in separable form unlike \eq{eq:add_emm_los}.  
A similar language was previously applied to the modeling of the SZ effect, also highlighting this aspect \citep{Chluba2012moments}.
One can now think of these moment maps, $\eta(\hat{n})$, as direct astrophysical observables. The simplest first order moments inform us about how the parameter along various lines of sight differ from the pivot parameter, while the second order moments inform us about the variance in the emission characteristics and so on.

In the final step, the generalized foreground modelling presented in \eq{eq:mom_sed_mod} is cast into a form that can be easily incorporated into the ILC machinery, yielding the \milc approach. By merely introducing the various SED derivatives as spectral constraints (as discussed in \sec{sec:ilc}) one can then simultaneously solve for the foreground moments maps as well as the cosmological observables like the CMB temperature, $y$ and $\mu$ distortion maps as we present in the following sections.
%

Some of the benefits of this approach for anisotropic signals were recently illustrated for the relativistic SZ temperature mapping \citep{Remazeilles2019} and the extraction of $\mu T$-correlations \citep{Remazeilles2018mu}. 
In this work, we unveil the real potential of this method, extending it to the extraction of tiny monopolar signals and also demonstrate that introducing many moment constraints can ensure an unbiased recovery of components while maintaining and even improving uncertainties. We note that the standard ILC and constrained ILC are special cases of {\milc}.

\vspace{-5mm}
\section{Simulations}
\label{sec:sims}
In this section, we summarize the details of our simulations. We assume a wide frequency coverage of $30-3000$ GHz, with 30 logarithmically-spaced channels and assume that the measurements are absolutely calibrated. Our multi-frequency simulated maps consists of the following components: blackbody, spectral distortion components $y$ and $\mu$ that characterize the deviation from the blackbody, foreground contamination in the maps and measurement noise. Below we provide specific details of how each of these components are injected in our simulations.

\textit{Blackbody:} In our simulations we include the CMB monopole and its anisotropies $\Delta_T(\hat{n}) = \Delta T (\hat{n}) / T_0$, characterized by the fiducial CMB power spectrum. The CMB SED is given by
\beq
I^{\rm CMB}_{\nu} (\hat{n}) = \left[ \Delta^0_T + \Delta_T(\hat{n})  \right] \frac{\partial B_{\nu}}{\partial T} T \Big\vert _{T_0}  \,,
\eeq
where $B_{\nu}$ denotes the Planck function. We assume that the electromagnetic spectrum of these fluctuations is described by a blackbody at a temperature $T_{0}=2.7255 ~\rm K$\footnote{\update{Since we assume that the temperature of 2.7255 K has been subtracted from the data, only $\Delta^0_T$, the correction to this temperature, is included in our simulations.}}, set to the currently best known measurement \citep{Fixsen1996, Fixsen2011}.
Note that we do not assume the CMB temperature to be known to arbitrary precision and solve for the monopole temperature $\Delta_T^0$ which in our simulations is set to a value of $10^{-4}$, about one order of magnitude below the error on the current CMB monopole temperature measurement. A temperature anisotropy map is constructed assuming a power spectrum for the fiducial cosmological model. When analyzing the simulations, we assume no prior knowledge about anisotropies of the CMB.

\textit{y-distortions: } These distortions are sourced via inverse Compton scattering of CMB photons off energetic electrons. The dominant contribution to this signal is sourced at low redshifts, by scattering of hot electron gas inside galaxy clusters \citep{Zeldovich1969, Mroczkowski2019}. To inject $y$-distortions in our simulations, we use the Compton $y$-map of \citet{Sehgal2009} available at \href{https://lambda.gsfc.nasa.gov/toolbox/tb_sim_ov.cfm}{LAMBDA-simulations}. We only use the non-relativistic thermal SZ spectrum for this work, as given by
\beq
I_{\nu}^{y}(\hat{n}) = y(\hat{n}) \left[x~ {\rm coth}\left( \frac{x}{2}\right)-4\right] \frac{\partial B_{\nu}}{\partial T} T \Big\vert _{T_0} \,,
\eeq
where $x= h \nu / k_B T$. We ignore the relativistic corrections to the average SZ spectrum \citep{Hill2015}, however, for detailed forecasts of the distortion sensitivities this should be included. The Compton-$y$ parameter field is a positive and non-Gaussian field and hence has non-zero positive monopole. The ensemble averaged $y$ parameter has been predicted to be $\langle y(\hat{n})\rangle = \bar{y}  \simeq 2 \times 10^{-6}$  in the fiducial cosmological model \citep[e.g.,][]{Refregier2000, Hill2015}. On a chosen patch, the sky averaged mean $\bar{y}$ may not necessarily match the ensemble-averaged mean. In our simulations we thus ensure that the sky-averaged value is close to this expected monopole, by directly adding/subtracting a monopole component to the $y$-map.

\textit{$\mu-$ distortions:} The dissipation of acoustic modes in the early universe ($z > 50,000$), introduces a $\mu$-distortion in the CMB radiation \citep{Sunyaev1970diss, Hu1993, Chluba2012}. While this signal also has spatial fluctuations, these are usually significantly smaller than the monopole distortions, in complete analogy with the CMB monopole temperature and hundreds of micro Kelvin fluctuations as a function of direction\footnote{More significant $\mu$-distortion anisotropies can be caused by primordial non-Gaussianity \citep[e.g.,][]{Pajer2012, Ganc2012, Razi2015, Ravenni2017}.}. In this work we thus only include the spectral distortion to the monopole spectrum in our simulations, while ignoring spatial fluctuations. The spectrum of the $\mu$-distortions is given by,
\beq
I_{\nu}^{\mu} = \mu \left[ \frac{1}{2.1923} - \frac{1}{x}\right]\frac{\partial B_{\nu}}{\partial T} T \Big\vert _{T_0} \,,
\eeq
where $\mu$ denotes the amplitude of the monopole signal and all the symbols have the same meaning as before.
Unlike the $y$-distortions, for which we have measurements from {\it Planck}, we only have upper limits on the $\mu$ distortions and upcoming experiments will aim at improving these limits and potentially making a detection. For this demonstration study we inject a $\mu$-distortion monopole signal with an amplitude of $\mu = 10^{-6}$, roughly two orders of magnitude below the current best limit set by {\it COBE/FIRAS} \citep{Fixsen1996}.

\textit{Foregrounds:} For our demonstration, we considered the following foreground components: synchrotron, free-free and thermal dust in our simulations, although we focus our discussion mainly on dust-only simulations. We generate the foreground skies using the Planck Sky Model (PySM) \citep{Thorne2016}. In particular we use the "d2" model for thermal dust, "s2" model for the synchrotron and the nominal free-free model in our simulations. The details of these component models can be found in the original PySM paper. 

Finally, we include some simple instrumental effects in our simulations: beam smoothing and measurement noise. For the instrumental beam we assume a uniform FWHM=30 arcminutes for all the channels. We take care to carry out the smoothing operation after duly converting the anisotropy maps returned by PySM from Rayleigh-Jeans temperature units to intensity units of Jy/sr. This small detail is important when we want to properly propagate the moments generated from this smoothing operation\footnote{\update{This aspect is not important when translating between antenna temperature to Jy/sr. However, the conversion between the commonly used $\mu K_{\rm CMB}$ units and intensity, strictly speaking, is a non-linear operation as it involves the inversion of the Planck function and in this case the smoothing and unit conversion operations do not commute, a detail that is potentially important.}}. 

We assume a constant noise RMS for all the channels, however, we do carry out the analysis for different \update{values: [5000,500,50,5] Jy/pixel. In our simulations the pixel size is 13.74 arcminutes. We use this information to translate these noise RMS values into all-sky sensitivities, yielding: [5.6, 0.56, 0.056, 0.0056] Jy/sr, respectively. Note that while 5 Jy/sr is representative of all-sky {\it PIXIE} sensitivity \citep{Kogut2011, Kogut2016SPIE}, 0.56 Jy/sr is representative of {\it SuperPIXIE} \citep{Kogut2019WP} and finally 0.056 Jy/sr is a little better than the full sky sensitivity for {\it Voyage~2050} \citep{Chluba2019Voyage}. Gains in angular resolution may be conceivable with concepts like {\it Millimetron}\footnote{\url{www.millimetron.ru/}} or by combining traditional CMB imagers with spectrometers \citep{PRISM2013WPII, Jacques2019Voyage}.}
A more detailed study covering more complex sky-models (e.g., including AME, CIB and CO emission) and instrumental aspects \update{(such as optimization of angular resolution and spectral parameters)} will be carried out elsewhere \citep{Rotti2020inprep}.

\subsection{Analysis strategy}
\label{sec:ana_strat}
For speedy analysis we evaluate the \milc algorithm on flat sky patches, however, this method is easily extended to the full sky analysis. All our simulations are generated at the {\tt Healpix}  \citep{Healpix} resolution of ${\rm NSIDE}=256$. From each of the frequency maps, we then extract sky tiles of dimensions $25^{\circ} \times 25^{\circ}$ centered on the galactic coordinate $(\ell,b) = (0^{\circ}, 30^{\circ})$, using the {\tt Healpix} gnomonic projection functionality. \update{We emphasize that identical analyses were carried out on different sky locations and we found qualitatively similar results, however here we choose to present results at this randomly chosen sky location.}

\update{
Our \milc filters are implemented in Fourier space and specifically the covariance is estimated as,
\begin{equation}
C^{k_{\rm bin}}_{\nu \nu'} = \frac{1}{(2 \pi)^2} \int_{k_{\rm bin}} \tilde{d}_{\nu} (\vec{k})  \tilde{d}_{\nu'} (\vec{k}) d^2 k\,,
\end{equation}
where  $\tilde{d}_{\nu} (\vec{k})$ denotes the Fourier coefficient of the multi-frequency data. Note that we do not remove the monopole term from the Fourier coefficients when making the binned frequency-frequency correlation matrix estimate\footnote{\update{It is important to bear in mind this detail when comparing the statistics estimated from the recovered component maps to those analytically estimated.
}}.}

When constructing the moment SED vectors, one has to make a choice for the pivot parameters $\bar{\bf p}$ and pivot frequencies $\nu_0$. In principle the solution to the pivot parameter changes for the simulations with different sensitivities and frequency coverage, however, in favour of simplicity we choose the pivot parameter from the highest sensitivity simulations and thereafter keep them fixed. Furthermore, the solution to the pivot parameters also changes when including higher order moment SEDs to fit the sky SED and in principle would require updates, however, this again is not essential. In our analysis, to find reasonable pivot parameters we fit the base SED to the sky-averaged spectrum and we choose the pivot frequency $\nu_0$ such that it coincides with the peak of the dust emission. After this initial step, $\bar{\bf p}$ is fixed and used to add higher order moments. It is important to note that the construction of moment SEDs does not rely on knowing precisely the parameters that characterize the foregrounds, on the contrary, the moment maps($\eta$) are part of the solution and these inform us about the statistical properties of parameters characterizing the foregrounds. 

For speedy analysis we evaluate the \milc algorithm on flat sky patches, however, this method is easily extended to the full sky analysis. All our simulations are generated at the {\tt Healpix}  \citep{Healpix} resolution of ${\rm NSIDE}=256$. From each of the frequency maps, we then extract sky tiles of dimensions $25^{\circ} \times 25^{\circ}$ centered on the galactic coordinate $(\ell,b) = (0^{\circ}, 30^{\circ})$, using the {\tt Healpix} gnomonic projection functionality. \update{We emphasize that identical analyses were carried out on different sky locations and we found qualitatively similar results, however here we choose to present results at this randomly chosen sky location.}

The multi-frequency simulated data, along with a subset of SEDs, are passed to the \milc algorithm, which then returns a set of component separated maps with a one to one correspondence to the SEDs. At the first stage of the analysis, we progressively increase the number of signal SEDs (i.e., CMB, $y$ and $\mu$) passed to \milc, which then returns the respective signal component maps. At the second stage, in addition to the signal SEDs we also pass the moment SEDs, progressively increasing the number of foreground vectors that are passed to the algorithm. \milc then returns all the signal maps along with the moment maps. In summary, the very first iteration of \milc only solves for the map of the CMB and at the final iteration the algorithm returns maps of all the signals, $[\Delta_T , y ,\mu]$, and multiple foreground moment maps, $\eta$. This analysis procedure is repeated on each set of multi-frequency simulations, where the only variable is the level of noise in the maps.
\vspace{-0mm}
\section{Results}
\label{sec:results}
In this section, we discuss the results of applying the \milc to the simulated data. We restrict the presentation and most of the discussion to results from analysis on simulations that only include thermal dust emission as foregrounds, since our primary aim is to demonstrate that \milc can not only be used to make robust (bias- and noise-reduced) measurements of anisotropic signals but also to extract tiny average-sky (monopolar) signals.
\begin{figure}
\hspace*{-0.18cm} 
\includegraphics[width=\columnwidth]{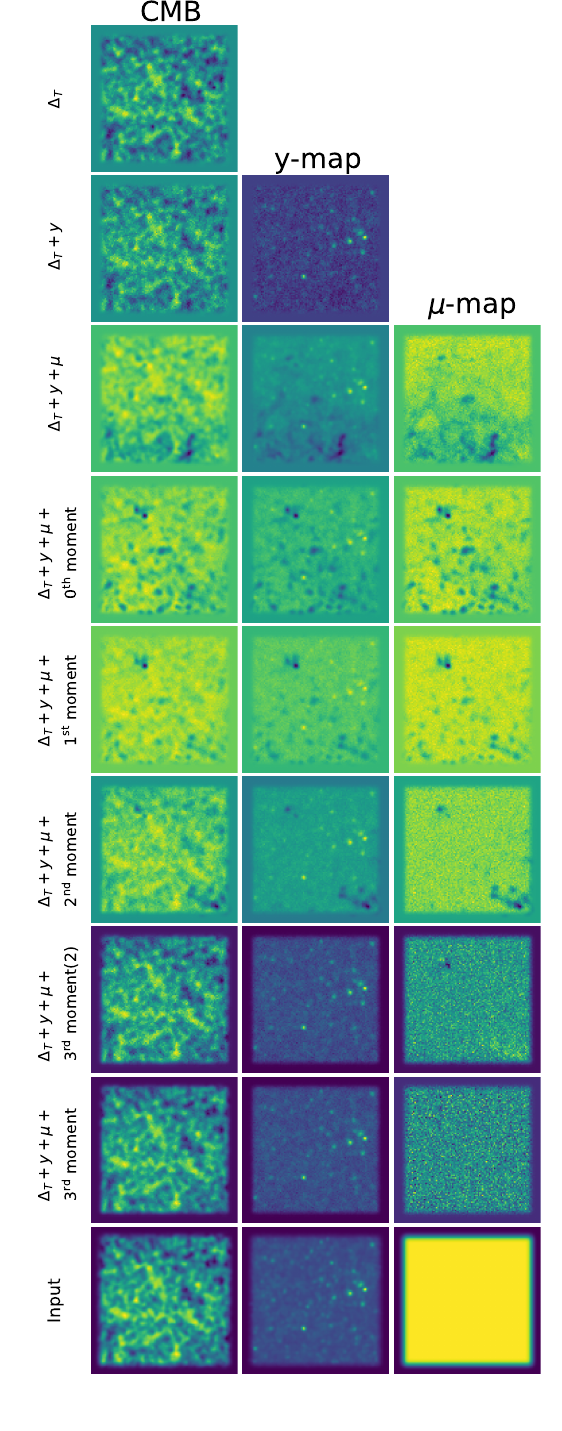}
\caption{This figure depicts the component-separated signal maps returned by the \milc algorithm, for a varying number of SEDs passed to it. \update{The annotations to the left of the image detail the set of SED vectors passed to \milc. These solutions are found for simulations with 50 Jy/px noise RMS. The input component maps are shown in the last row.}}
\label{fig:sig_rec_fn_of_mom}
\end{figure}

The thermal dust emission is characterized by a modified blackbody: $S^{\rm d}_{\nu} = \nu^{\alpha} B_{\nu}(1/\beta)$, where $\beta$ is the inverse of the dust temperature. Following \eq{eq:mom_sed_mod}, the moment expansion is performed by taking different order derivatives with respect to the parameters (${\bf p} = [\alpha, \beta]$) characterizing the SED. The SED vectors characterizing the different moments are simply given by $S^{ij}_{\nu} = \mathcal{N}_{ij} \partial_{\alpha}^i \partial_{\beta}^j S_{\nu}$, where\footnote{This normalization only changes the overall amplitude of the moment SED and not its shape and hence it is only important when we care about the amplitude of the moment maps. Given some normalization there is no ambiguity on the amplitude of the moment maps.} $\mathcal{N}_{ij}$ is some normalization and the moment order is given by: $o = i + j$. The moment maps corresponding to each of these moment SED vectors are denoted by $\eta_{\alpha^i  \beta^j}$. Note that $\eta_{\alpha^0  \beta^0}$ refers to the amplitude of dust emission and is denoted by $A_{\rm D}$. We include a maximum of 10 moment SED vectors in our analysis which corresponds to maximum moment order of $o=3$.

\begin{figure}
\includegraphics[width=1.\columnwidth]{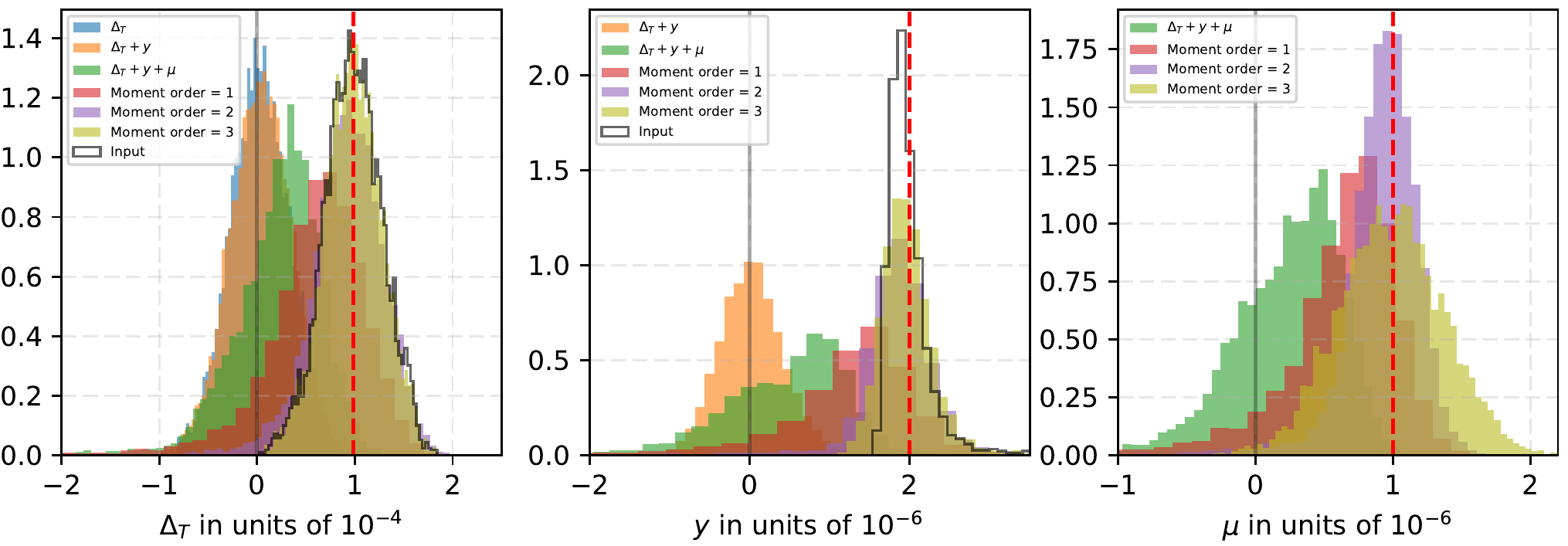}
\caption{This figure depicts the evolution of the 1PPDF of the estimated component maps from analysis performed on simulations with a noise RMS = 50 Jy/px. The dashed line indicates the value of the monopole estimated directly from the input component maps.}
\label{fig:1ppdf}
\end{figure}

\begin{figure}
\includegraphics[width=\columnwidth]{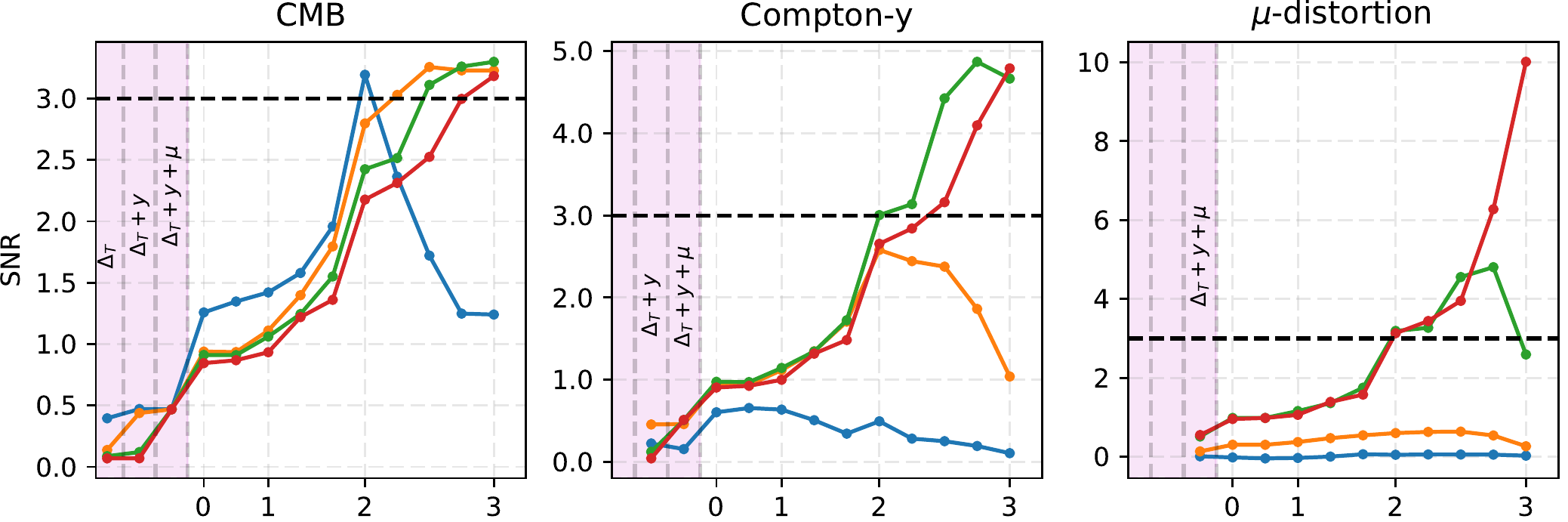}
\includegraphics[width=\columnwidth]{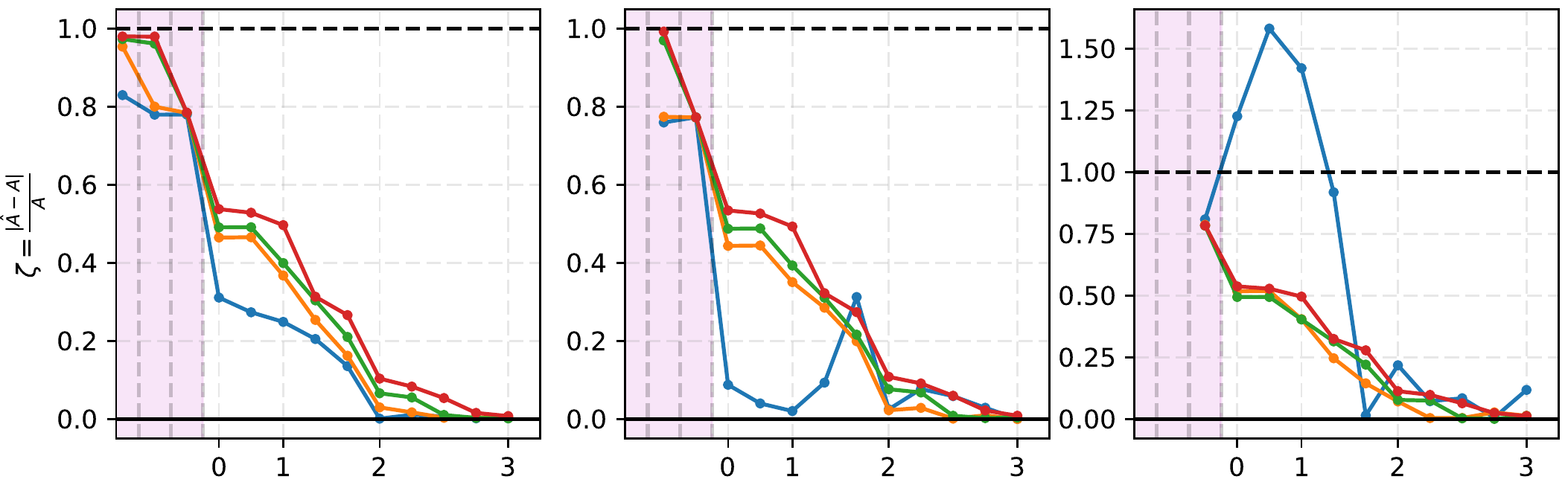}
\includegraphics[width=\columnwidth]{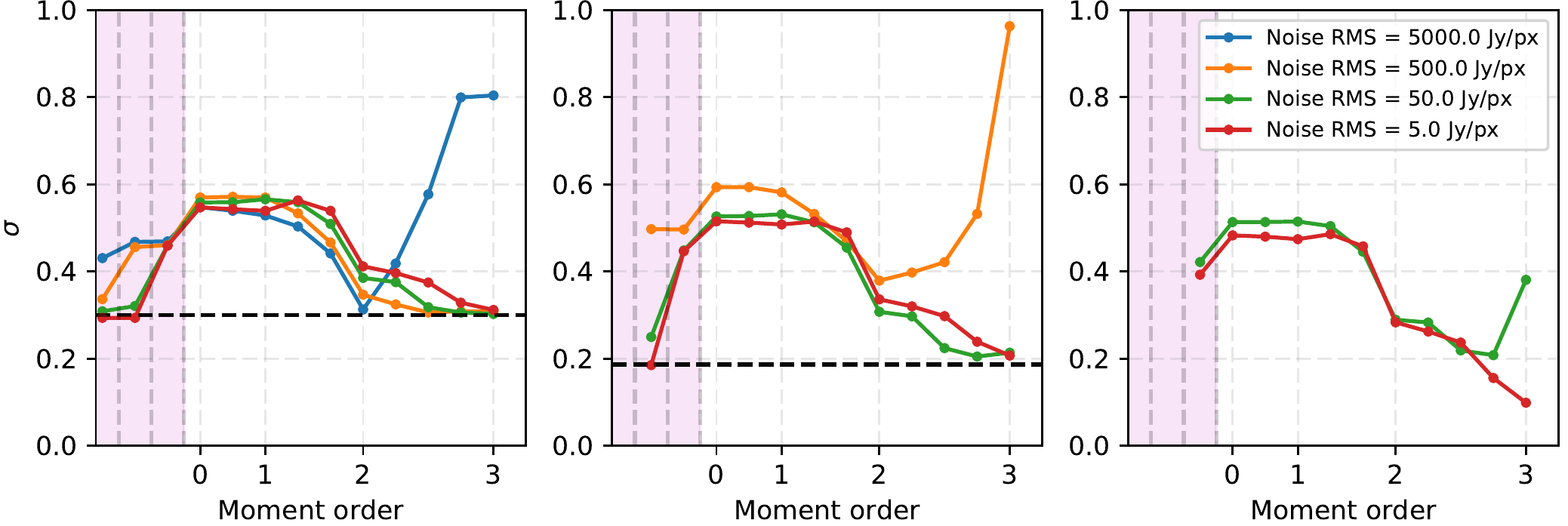}
\caption{Evolution of the 1- and 2-point statics of the recovered component maps, for varying number of vectors passed to the \milc at different noise-levels. The top panels depict the evolution of the detection significance of the estimated monopole amplitude. \update{The middle panels depict evolution of the estimated bias($\zeta$).} The bottom panels depict the evolution of the standard deviation of the map in units of the true monopole of the respective component map. Here, the dashed black line marks the standard deviation of the input component map.}
\label{fig:map_stat}
\end{figure}

\begin{figure}
\includegraphics[width=1\columnwidth]{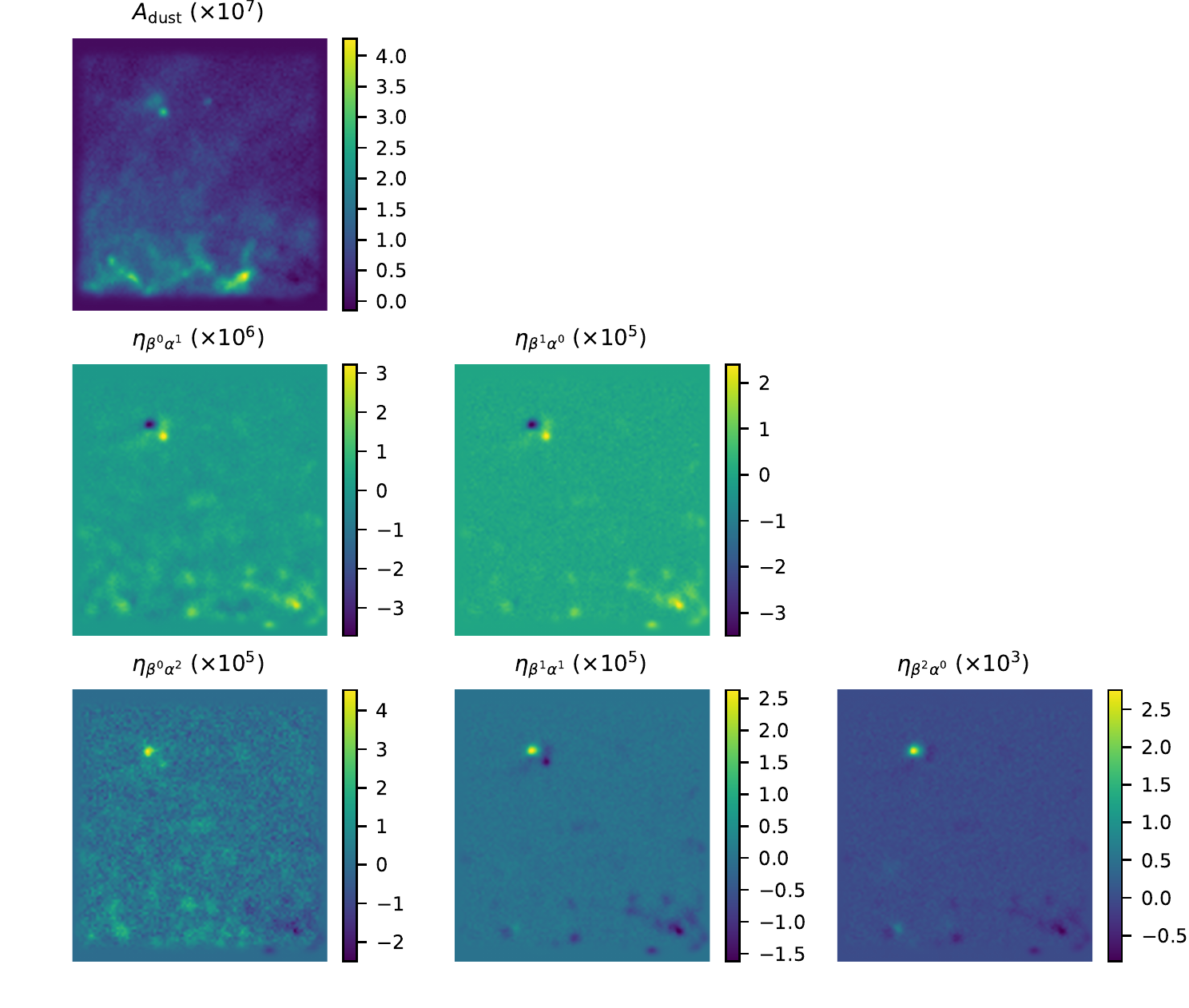}
\caption{This panel of images depicts the dust foreground moments up to second order from analysis on simulations with a noise of 5 Jy/px. Higher order moment maps are correlated but in detail exhibit slightly different spatial morphologies.}
\label{fig:moment_maps}\end{figure}

Figure~\ref{fig:sig_rec_fn_of_mom} depicts the component separated signal maps returned at each evaluation of \milc on simulations that include noise at the level of 50 Jy/pixel. Recall that our simulations are absolutely-calibrated and hence retain information about the monopoles in the respective component maps (unlike experiments like \WMAP and \Planck, which carried out differential measurements). To diagnose the measurement of the monopole signal we study the one-point probability distribution function (1PPDF) of the recovered component maps. Figure~\ref{fig:1ppdf} depicts the 1PPDF of a subset of these maps. Figure~\ref{fig:map_stat} depicts the evolution of the 1 point and 2 point statistics of maps, for varying number of vectors passed to \milc and for different levels of noise in the simulations. Figure~\ref{fig:moment_maps} depicts the first 6 foreground moments (i.e., up to second order for dust) estimated on simulation with the lowest noise. With the help of the visual aid provided by these figures we now discuss the salient features of solutions delivered by different iterations of the \milc.

The first row of \fig{fig:sig_rec_fn_of_mom} depicts a map of the recovered CMB, and this iteration of the \milc corresponds to the standard ILC, which only projects out the CMB and is completely blind to the presence of other components in the map, subsuming these into the variance minimization process. Comparing to the true CMB anisotropy map (right column of \fig{fig:sig_rec_fn_of_mom}) one notes that many of the features in the true CMB map are indeed recovered, however, there are residuals, particularly noticeable as negative holes that coincide with the position of the brightest galaxy clusters also seen in the injected $y$-map. We also note from \fig{fig:1ppdf} and \fig{fig:map_stat} that the mean is consistent with zero, in spite of a non-zero monopole in the simulation. This recovered map of CMB anisotropies is also significantly noisier and this can be understood (in hindsight) as the cost of not projecting out other components in the map.
The second row in \fig{fig:sig_rec_fn_of_mom} depicts the simultaneous separation of the CMB and the Compton-$y$ map, constructed by mutually de-projecting the respective SEDs. This iteration of the \milc now correspond to the standard cILC, which is one of the methods used by the \Planck collaboration. Note that this achieves a reasonable recovery of both the CMB and the Compton-$y$ map. The SZ residuals in the CMB maps, seen in the previous iteration of the \milc, have now disappeared. There are more subtle biases in these maps that are not as visible in the considered simulation but become more prominent at reduced experimental sensitivity. The recovered component-separated maps continue to have excess noise, which results from not de-projecting other map components, namely, foregrounds. The means of both the recovered fields are still consistent with zero as seen in \fig{fig:1ppdf} and~\ref{fig:map_stat}.

The third row in  \fig{fig:sig_rec_fn_of_mom} depicts the simultaneous separation of the CMB, $y$ and the $\mu$ distortion map, constructed by mutually de-projecting the respective SEDs. This iteration of the \milc corresponds to the cILC, with three different but well-known SEDs. \update{Many of the large-scale features in the CMB, and $y$-maps at the location of some of the brightest galaxy clusters are still recovered.} However, the recovery of the component maps has severely degraded as compared to the previous two iterations of \milc. This is also apparent from the increase in the variance of the CMB and $y$ maps as seen in \fig{fig:map_stat}. It is interesting to note that the peak of the 1PPDF for the CMB and $y$-parameter maps start showing a shift towards the input monopole values as seen in \fig{fig:1ppdf}.

The subsequent rows of \fig{fig:sig_rec_fn_of_mom} depict the simultaneous separation of the signal components, when projecting out an increasing number of foreground moment vectors (1 in the $4^{\rm rd}$ row, 3 in the $5^{\rm th}$, 6 in the $6^{\rm th}$, 8 in the $7^{\rm th}$ and finally 10 in the $8^{\rm th}$ row). 
With these iterations we are truly entering a new regime, where foreground-averaging effects are gradually being captured. In fact the foreground along different lines of sight cannot be characterized by a single SED, this is in stark contrast with the previous iterations of the \milc, where all the passed signal SEDs nearly perfectly characterize the respective components along different lines of sight. 
\update{The recovered component maps depicted in rows 4-6 show a slow reduction in the residuals and then in rows 7 and 8 the recovery of the components maps is nearly perfect and one cannot discern the differences between the maps.} This already shows that given sufficient frequency coverage and sensitivity, addition of moment terms does not come with unavoidable penalties in the recovery of signals but has a more complicated behaviour.
We can also use the variance of the maps for a joint assessment of the quality of the recovered components. Studying the bottom panel of \fig{fig:map_stat} reveals a much clearer picture of the evolution: the variance of the map increases until one includes the $0^{\rm th}$ order vector, thereafter every additional foreground vector passed to the \milc results in an improved recovery, reaching near ideal recovery once all foreground moment vectors up to the second order moment are projected out. Even in rows 7 \& 8, where it is hardly possible to discern map-level differences, the variance statistics indicates that there is continual improvement. It is also very important and exciting to note that on including these higher order moments the peak of the 1PPDF of the maps converges to the true monopole amplitude of the respective components as seen in \fig{fig:1ppdf}.

\update{We note that a statistically significant detection of the monopole amplitude only becomes possible on nearly including all the second order moment SED vectors and higher, however this trend also depends on the sensitivity of the measurements as seen in the top row of Fig.~\ref{fig:map_stat}. For CMB and y fields, which are inherently anisotropic, the $\sigma$ estimated from the map receives significant contribution from these intrinsic anisotropies. This is also reflected in the saturation in the SNR plots for these respective components. When measuring the monopoles of the CMB and $y$ fields, it will be important to accurately model and subtract the error arising from the intrinsic variance of these fields, in order to fairly assess the statistical significance of these detections (in this sense our assessments of the SNR for these fields is biased low). On the other hand, since our simulations included no anisotropic $\mu$ distortions, even though the $\mu$ distortions signal is smaller it is detected at a higher statistical significance for simulations with the lowest noise. The CMB and $y$ signals being intrinsically larger is reflected in the observation that a statistically significant detection of the monopole amplitudes of these fields is possible even at lower sensitivities.}

The details of the evolution of these statistics with the number of moment vectors passed to the \milc is of course not generic and depends on the sensitivity and frequency coverage of the experiment and the foregrounds model that is injected. \update{Further these details also depend on how the data correlation matrix is defined, however a discussion on these details is beyond the scope of this work and will be discussed in \citet{Rotti2020inprep}.} It will be superfluous and redundant to show the maps and 1PPDFs for analysis carried out on each simulation set with different noise. The evolution of the detection significance of the monopole amplitude, its bias and variance of the recovered component maps from analysis on simulations with different noise are neatly summarised in \fig{fig:map_stat}.

Here we would like to highlight one generic trend that in the first four iterations of the \milc (these are methods employed for various work in current literature), the variance of the recovered component map {\it increases} when adding constraints, as seen from the bottom panel of \fig{fig:map_stat}. This gives rise to the misleading expectation that adding constraints invariably comes with a noise penalty. However, as the bottom panels of \fig{fig:map_stat} demonstrate, this is a faulty extrapolation. We generically find that after projecting out an optimal number of foreground moment vectors, the variance of the recovered component map reaches a minimum, which is below or comparable to the error when these de-projections are not carried out (see \fig{fig:map_stat}).

It is also very interesting to note that, when considering the highest sensitivity (RMS=5 Jy/px, red line in Fig.~\ref{fig:map_stat}), we find the simple ILC or cILC (only CMB and CMB+$y$, respectively) do perform {\it on par} with the \milc, indicating that signal SED orthogonality is a direct function of sensitivity. However, for all our lower sensitivity simulations, the \milc with inclusion of higher order foreground moments has the best performance.
In hindsight, this behaviour is easy to understand: when one does not project out foregrounds, these are included in the noise budget of the recovered component maps. By projecting out an `optimal number' of foreground components, the noise budget of the component separated maps is reduced. The `optimal number' of foreground components can be assessed by studying when the variance of the component maps begins to increase after reaching the minimum (see the blue line for CMB variance and orange for $y$-map variance in bottom panels of \fig{fig:map_stat}). To be able to make these projections, one requires enough channels, and this diagnostic could be used to optimize future CMB missions.

The foreground moment maps are themselves astrophysical observables and these can be used to gain deeper understanding of the galactic foregrounds. These moment maps are naturally returned by the \milc and an example set of the first measurement of foreground moment maps on simulations and are shown in \fig{fig:moment_maps}. These moment maps are related to the spectral parameters characterizing the IGM, but elucidating this connection requires more work. Current and future analysis on multi-frequency microwave data will deliver these observables as by-products.
%

\vspace{-0mm}
\section{Discussions and conclusions}
\label{sec:discussions}
We have demonstrated that the scope of ILC approaches can be significantly extended by using the language of moments, resulting in the \milc method introduced here (Sect.~\ref{sec:methods}).
Using sky simulations, we have demonstrated that the \milc can not only be used to improve the recovery of anisotropic components but also allows an extraction of monopole signals. This opens new ways of thinking about mitigating foreground challenges that are essential to future CMB spectral distortion studies.

Specifically, we have demonstrated that it is important to consider higher order moments into account, without which both the monopole and the anisotropy measurements can be significantly biased (see \fig{fig:sig_rec_fn_of_mom} and \fig{fig:1ppdf}). Furthermore, we have clarified that increasing the number of moments does not generically lead to more noise in the recovered component maps (see \fig{fig:map_stat}). On the contrary, we have demonstrated that projecting out an `optimal number' of foreground moments will invariably lead to a more robust and less noisy recovery of the component maps. 

Our study also shows that the \milc can have more optimal performance as compared to simple ILC and cILC methods. Only for very high sensitivity measurements are the performances of the ILC and cILC on par with \milc. These lead to a revision in the understanding of how to apply ILC methods. 
The variance of the recovered component maps can be used to estimate the `optimal number' of foreground vectors to reach best performance. This number can be directly translated into the requirement of the number of frequency channels at some given sensitivity and has the potential of being a quantitative metric for designing future CMB experiments (under some assumptions about the sky foregrounds).

While the discussion presented in this work is focused on simulations which include only thermal dust as foregrounds, we have also confirmed that this method works when all the foregrounds (synchrotron, free-free and dust) are included. This turns into a high-dimensional problem, and one has to worry about the most relevant SED vectors and the order in which they are included. This is an ongoing study \citep{Rotti2020inprep}, the details of which are beyond the basic idea being presented in this work.
%
The foreground moment maps are a natural by-product of \milc and we foresee these becoming an important measurable with future, high-sensitivity multi-frequency microwave measurements. \milc is currently being applied to \Planck maps and the measurement of foreground moment maps from this analysis will be reported in a future publication.
However, several questions call for further investigation:

\JCparagraph{Effect of un-accounted signals:} The work presented here assumes that the base SEDs, characterizing the all the foregrounds, are known and can be supplied to the \milc. This allows \milc to capture all the effects from averaging processes. We have shown that \textit{not} de-projecting the higher order moments can come at the cost of excess noise and biases in the recovered component maps. This aspect will become even more apparent as we work with higher sensitivity data to target low signal to noise components like for CMB spectral distortions and $B$-modes. Encountering foregrounds with unmodelled SEDs will lead to a poorer component separation. However, having even an approximate SED for these components can improve the signal recovery. Applied to real data, \milc should thus be thought of as a {\it semi-blind} component separation method.
Adding Synchrotron, CIB and free-free moments is straightforward \citep{Chluba2017foregrounds}, however, developing a moment description for some of the other known foregrounds (e.g., anamolous microwave emission, extra-galactic CO) will be an important step in generalizing \milc. 

\JCparagraph{Low multipole leakage to monopole and sky-coverage:}  In this work we carried out the analysis on a flat-sky patch and implicitly assumed that local patch monopole is the same as that of the global monopole. For the CMB and $y$, which do have significant amount of anisotropies, the local patch monopole can differ significantly from the global monopole owing to contributions from large scale ($\lambda \geq$ patch size) fluctuations. For this reason the local patch monopole measurement of these components could be noticeably biased. However, in standard cosmology the $\mu$-distortions are expected to be monopolar, and hence for this component, the amplitude of the local monopole is expected to be the same as the global monopole. This suggest that if one wants to measure $\mu$ distortion monopole, making deep measurements on a flat-sky patch could be a viable strategy. On the other hand, an ultimate test of isotropy of the primordial distortion signal \citep[e.g., also including the cosmological recombination lines ][]{Sunyaev2009, Chluba2016CosmoSpec} will require all-sky measurements.
A detailed study with the goal to optimize for CMB spectral distortion science goals will be presented in a future publication.

\JCparagraph{Truncation and ordering of vectors:} The  variance of the recovered component maps was seen to have a non-monotonic behaviour, reaching a minimum at some optimal number of moment vectors (see Fig.~\ref{fig:map_stat}). It will be important to study if the moment ordering could be altered to obtain a monotonic reduction in variance of the component separated maps. This will help in devising the optimal strategies for taming this high dimensional optimization problem. Additional benefits could stem from orthogonalizing moment SEDs (e.g., Gram-Schmidt or principle component analysis) prior to the \milc analysis. This will enable us to more clearly rank the various moment vectors according to their expected signal to noise levels. Ultimately, a comparison of the target signal level (i.e., $\mu$) to the level of a moment SEDs has to be used to define truncation criteria for the analysis. Other diagnostics, based on Bayesian evidence, Shannon entropy or $\chi^2$-gains, could also be used for this purpose.

\JCparagraph{Optimality for monopole recovery:} \milc performs a pixel by pixel recovery of the signal field. While this approach has been (empirically) proven to be `optimal' for anisotropic signal, it is not immediately obvious that it is optimal for the recovery of monopole signals. With single pixel sky-averaged SED measurements, one can make significant gains ($\simeq \sqrt{N_{\rm pix}}$) in sensitivity, however, possibly at the cost of increasing the foreground complexity due to generation of moments, a natural consequence of the averaging process. In principle this questions is relevant to measurements of all the long wavelength modes on the sky (i.e., low-$\ell$ CMB anisotropies). Working with full resolution maps, will invariably lead to lower sensitivity per pixel but provides access to additional information about pixel-pixel correlations for different foreground components and, possibly, a lower foreground complexity in each pixel. It can be anticipated that the answer lies somewhere in the middle. It will thus be essential to compare and contrast the performance of these two analysis approaches, to understand how to best combine them for extraction of spectral distortion signals.

\JCparagraph{Combination of datasets:} 
High-sensitivity and high-resolution anisotropy measurements of the CMB will soon become available. The \milc method makes it clear how we can blend CMB anisotropy measurement to help with cleaning of the absolutely-calibrated measurements planned for the future \citep[e.g.,][]{Kogut2019WP,Chluba2019Voyage}. Increased frequency coverage at low frequencies ($\nu\lesssim 10\,{\rm GHz}$) has also been shown to be important for the recovery of small $\mu$-distortion signals \citep{abitbol_pixie}. Again the \milc method can be extended to include external information from low-frequency data sets. This also means that a detailed study of how to propagate of systematic effects when combining various data sets will be required.

\JCparagraph{Application to other problems:} 
The \milc method is not limited to studies of primordial CMB distortion signals, but similarly can be applied to the extraction of the global 21cm signal from the cosmic dark ages \citep[e.g.,][]{Pritchard2008, Fialkov2018}. At low frequencies, the dominant foregrounds are due to radio sources and galactic synchrotron, which can both be modelled using a power-law moment expansion \citep{Chluba2017foregrounds}.
In a similar manner, it is straightforward to extend the \milc approach to CMB $B$-mode searches, ensuring that foreground residuals caused by averaging effects remain under control \citep{Remazeilles2020}. In this case, two moment hierarchies have to be introduced, demanding high sensitivity and broad spectral coverage. De-correlation effects across frequency can also be modelled in this way. Moment expansions at the power-spectrum level can provide further insight and leverage \citep{Mangilli2019}. 
We plan to explore these possibilities in future works. 

The component separated maps depicted in the third column of \fig{fig:sig_rec_fn_of_mom} represents a study analogous to that presented in \citep{Remazeilles2018mu}. We see that there are significant biases at this iteration of the \milc, hinting that one could make significant improvements in forecasts for $\mu$-$T$ correlations by including higher order moments
into the analysis. Of course these studies will need to be repeated with all the relevant foregrounds and the instrument configurations considered in \citep{Remazeilles2018mu}. The utility of \milc for the extraction of $y$-$T$ correlations to study primordial non-Gaussianty \citep[e.g.,][]{Ravenni2017} should also be carefully considered.

Finally, our study also indicates that even for the CMB temperature recovery, the simple ILCs and cILCs are not fully optimal and may have significant residuals. This could relate to some of the large-scale anomalies seen by \Planck and \WMAP (namely lack of  power and isotropy violation on large-angular scales). Revisiting this question using \milc on real data could thus prove highly instructive and may also inform us about expected foreground complexities for future $B$-mode searches, both questions we plan to address in the future.

\vspace{1mm}
{\small
{\noindent \it Acknowledgments:} This work was supported by the ERC Consolidator Grant {\it CMBSPEC} (No.~725456) as part of the European Union's Horizon 2020 research and innovation program.
JC was supported by the Royal Society as a Royal Society URF at the University of Manchester, UK. We thank Mathieu Remazeilles for useful discussions.
}



\bibliographystyle{mnras}
\bibliography{Lit,Lit1,mendeley_cmb} 


\end{document}